\documentclass{iau}
\usepackage[pdftex]{graphicx}
\usepackage{amsmath}

\title[Constraining X-ray-Induced Photoevaporation] %% give here short title %%
{Constraining X-ray-Induced Photoevaporation of Protoplanetary Disks Orbiting Low-Mass Stars}

\author[K. M. Punzi]   %% give here short author list %%
{Kristina M. Punzi$^1$, Joel H. Kastner$^1$, David Rodriguez$^2$, David A. Principe$^{3, 4}$, Laura Vican$^5$}

\affiliation{$^1$Laboratory for Multiwavelength Astrophysics, Rochester Institute of Technology\\
$^2$ Universidad de Chile\\
$^3$ N{\'u}cleo de Astronom{\'i}a de la Facultad de Ingenier{\'i}a, Universidad Diego Portales\\
$^4$ Millennium Nucleus Protoplanetary Disks\\
$^5$ University of California, Los Angeles}

\pubyear{2015}
\volume{314}  %% insert here IAU Symposium No.
\pagerange{119--126}
\jname{Young Stars \& Planets Near the Sun}
\editors{J. H. Kastner, B. Stelzer, \& S. A. Metchev, eds.}
\begin{document}

\maketitle

\begin{abstract}
Low-mass, pre-main sequence stars possess intense high-energy radiation fields as a result of their strong stellar magnetic activity.
This stellar UV and X-ray radiation may have a profound impact on the lifetimes of protoplanetary disks.
We aim to constrain the X-ray-induced photoevaporation rates of protoplanetary disks orbiting low-mass stars by analyzing serendipitous XMM-Newton and Chandra X-ray observations of candidate nearby (D $<$ 100 pc), young (age $<$ 100 Myr) M stars identified in the GALEX Nearby Young-Star Survey (GALNYSS).
\keywords{stars: low-mass, pre-main sequence; techniques: imaging spectroscopy; X-rays: stars}
\end{abstract}

\vspace{-0.8cm}
\section{Introduction}

Low-mass (M-type) stars represent some of the best targets for the discovery of potentially habitable exoplanets due to their low luminosities and the location of their habitable zones.
Presently, only a small number of planets have been detected around M stars, with terrestrial planets being common and giant planets being rare, although these results may be affected by a selection bias (e.g., \cite[Mulders et al. 2015]{Mulders2015}, \cite[Howard et al. 2012]{Howard2012}).
This trend may be a consequence of the intense high-energy radiation fields of low-mass, pre-main sequence stars.
A great deal of mass in protoplanetary disks is lost from the surface of the disk due to heating (photoevaporation) from high-energy radiation from the central star.
The X-rays from young stars drive disk dissipation and chemistry, influencing the timescale and conditions for exoplanet formation.
According to \cite[Owen et al. (2012)]{Owen2012}, stellar X-ray luminosity alone sets the photoevaporation rate.
However, \cite[Gorti et al. (2015)]{Gorti2015} demonstrate that X-ray spectral hardness is also important.
Hence, it is necessary to fully characterize the X-ray radiation fields incident on protoplanetary disks.

\vspace{-0.6cm}
\section{Serendipitously Detected X-ray Counterparts}

To constrain the X-ray-induced photoevaporation rates of protoplanetary disks orbiting low-mass stars, we can examine stars from the GALEX Nearby Young-Star Survey (GALNYSS; \cite[Rodriguez et al. 2013]{Rodriguez2013}) that have been serendipitously observed by either XMM-Newton or Chandra.
GALNYSS combines ultraviolet (GALEX) and near-IR (WISE and 2MASS) photometry with kinematics to identify candidate nearby (D $<$ 100 pc), young (age $<$ 100 Myr), low-mass (M-type) stars.
This survey has identified $>$2000 candidates, with most of the stars having spectral types in the range M3-M4.

\begin{figure}[hb!]
\begin{center}
\includegraphics[width=4in]{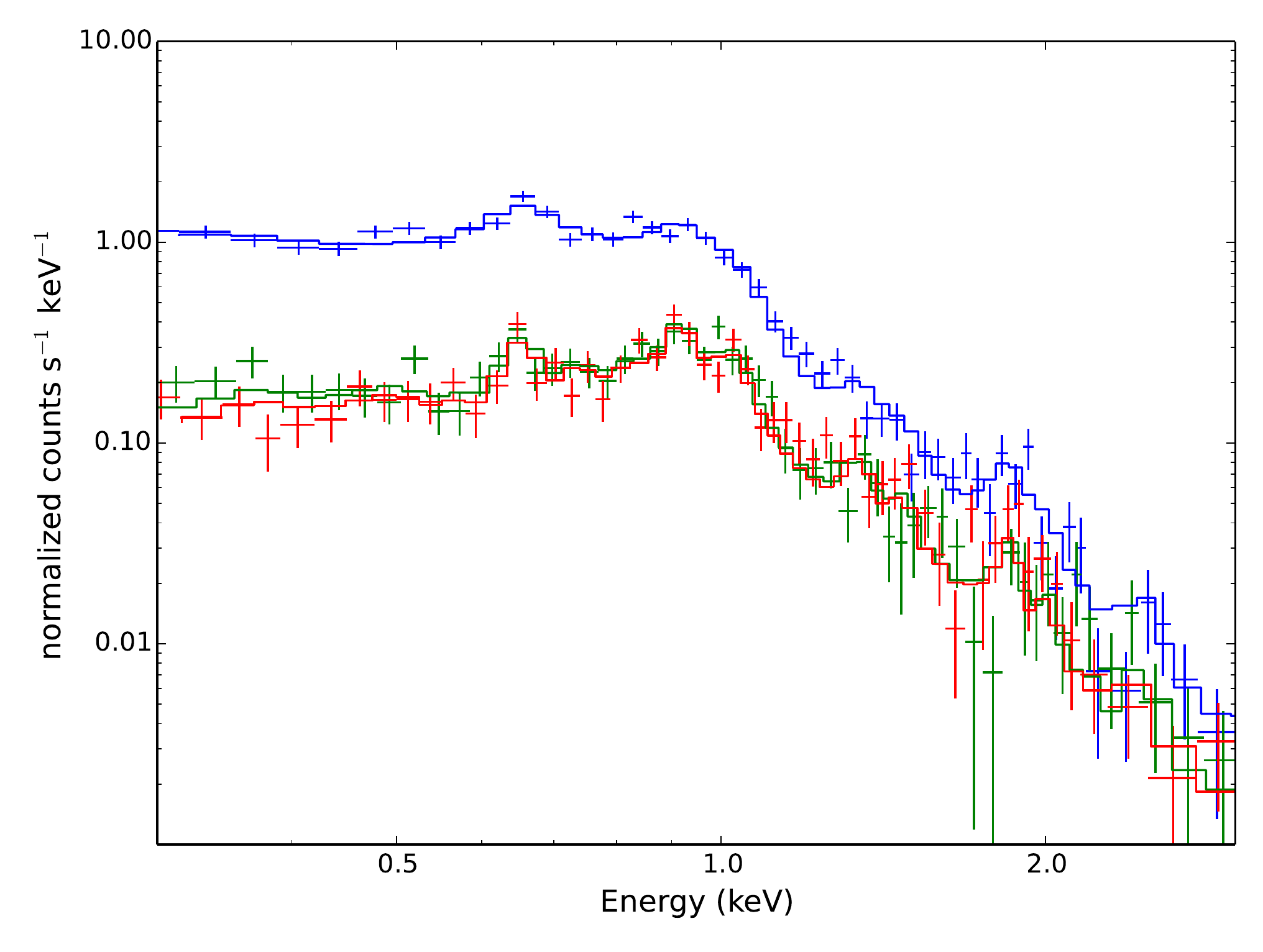} 
  \caption{XMM-Newton EPIC extracted spectra (crosses) of J061313.30-274205.6 (spectral type M3, $\beta$ Pictoris Moving Group candidate -- 99.24$\%$ likelihood) for pn (blue) and MOS (red and green) detectors. Overplotted are the best-fit models (histograms). Our model predicts plasma temperatures of $\sim3.6$ and $\sim12$ MK and an X-ray luminosity of $1.2\times10^{29}$ erg s$^{-1}$ at a distance of $\sim25$ pc.}
   \label{fig:spectrum}
\end{center}
\end{figure}

The XMM-Newton data for GALNYSS candidates are being reprocessed and analysed using the Scientific Analysis System following standard procedures \footnote{See the SAS documentation at http://xmm.esac.esa.int/sas/current/documentation/threads.}.
Spectra are extracted from the EPIC detectors using circular regions centered on the 2MASS/WISE positions of the objects.
Spectral modelling is performed with XSPEC.
An example of the extracted spectra and resulting model fit is displayed in Figure~\ref{fig:spectrum}.
Our models consist of X-ray spectra that result from optically thin plasmas in collisional ionization equilibrium (vapec model in XSPEC) \footnote{See the XSPEC documentation at https://heasarc.gsfc.nasa.gov/xanadu/xspec/manual/Models.html for a description of the models.}.
These emission spectra were combined with photoelectric absorption by using the XSPEC model wabs.

\vspace{-0.6cm}
\section{Conclusions and Future Work}

Our preliminary results demonstrate that serendipitous XMM-Newton observations of GALNYSS stars are capable of producing useful constraints on the stellar X-ray temperature and luminosity and hence, on photoevaporation due to X-ray irradiation.
These early results suggest that X-ray photoevaporation may not account for complete disk dispersal at ages $\sim$ 8-20 Myr.
For the entire sample of serendipitously observed GALNYSS stars, we will determine X-ray temperatures and luminosities, the potential correlation between M star L$_{X}$ and residual disk mass, and the mass dependence of L$_{X}$/L$_{bol}$.

\vspace{-0.5 cm}
\begin{thebibliography}{}

\bibitem[Gorti (2015)]{Gorti2015}
{Gorti, U., Hollenbach, D., \& Dullemond, C. P.}, 2015,
\textit{ApJ}, 804, 29.

\bibitem[Howard (2012)]{Howard2012}
{Howard, A. W., Marcy. G. W., Bryson, S. T., et al.} 2012,
\textit{ApJS}, 201, 15.

\bibitem[Mulders (2015)]{Mulders2015}
{Mulders, G. D., Pascucci, I., \& Apai, D.} 2015,
\textit{ApJ}, 798, 112.

\bibitem[Owen (2012)]{Owen2012}
{Owen, J. E., Clarke, C., J., \& Ercolano, B.}, 2012,
\textit{MNRAS}, 422, 1880.

\bibitem[Rodriguez (2013)]{Rodriguez2013}
{Rodriguez, D. R., Zuckerman, B., Kastner, J. H., et al.}, 2013,
\textit{ApJ}, 744, 101.

\end{thebibliography}
\end{document}